# Collective Behavior Induced Highly Sensitive Magneto-Optic Effect in 2D Inorganic Liquid Crystals

Tianshu Lan[1], Baofu Ding[1]*, Ziyang Huang[1], Fenggang Bian[2], Yikun Pan[1], Hui-Ming Cheng[1,3,4]* and Bilu Liu[1]*

[1] Shenzhen Geim Graphene Center, Tsinghua-Berkeley Shenzhen Institute and Institute of Materials Research, Tsinghua Shenzhen International Graduate School, Tsinghua University, Shenzhen 518055, China

[2] Shanghai Synchrotron Radiation Facility, Shanghai Advanced Research Institute, Chinese Academy of Sciences, Shanghai 201204, China

[3] Shenyang National Laboratory for Materials Science, Institute of Metal Research, Chinese Academy of Sciences, Shenyang 110016, China

[4] Advanced Technology Institute, University of Surrey, Guildford GU 27XH, UK

*Corresponding authors: ding.baofu@sz.tsinghua.edu.cn (B.D.); hmcheng@sz.tsinghua.edu.cn (H.M.C); bilu.liu@sz.tsinghua.edu.cn (B.L.)

**Abstract**

Collective behavior widely exists in nature, ranging from the macroscopic cloud of swallows to the microscopic cloud of colloidal particles. The behavior of an individual inside the collective is distinctive from its behavior alone, as it follows its neighbors. The introduction of such collective behavior in two-dimensional (2D) materials may offer new possibilities to achieve desired but unattained properties. Here, we report a highly sensitive magneto-optic effect and transmissive magneto-coloration via introducing collective behavior into magnetic 2D material dispersions. The increase of ionic strength in the dispersion enhances the collective behavior of colloidal particles, giving rise to a magneto-optic Cotton-Mouton coefficient up to 2700 $T^{-2}m^{-1}$ which is the highest value obtained so far, being three orders of magnitude larger than other known transparent media. We also reveal linearly dependence of magneto-coloration on the concentration and hydration radius of ions. Such linear dependence and the extremely large Cotton-Mouton coefficient cooperatively allow fabrication of giant magneto-birefringent devices for color-centered visual sensing.

**Introduction**

Liquid crystals (LCs) have been exploited in many applications including display, smart windows, light modulators, biomedicine and visually chemical and thermal sensors.[1, 2] Achieving orderly alignment of LCs in mild external stimuli, such as electric, magnetic, mechanical and thermal stimuli,[3-11] is crucial to fabricate high-sensitivity LC devices for a stimulus-optic response. In nature, one way to keep high order-formation for many creatures is to behave collectively, such as the clouds of swallows, bees and ants.[12-14] The behavior of an individual creature inside the collective is dependent on its neighbors, thereby being different from its behavior alone. Analogous to this macroscopic phenomenon, the microscopic colloidal particles have a similar collective behavior due to the interactions acting on the adjacent particles, including the repulsive electrostatic repulsion, van der Waals attraction and other inter-particle interactions.[15] It can be envisaged that the formed collective comprising orderly aligned individual colloidal particles which can serve as a functional unit and is expected to have higher sensitivity in response to the external stimulus than the individual counterpart.

Recently, thanks to its extremely large shape anisotropy and the associated magnetic/optical anisotropy, magnetic two-dimensional (2D) colloids have triggered intensive research activities, as their alignment can be tuned in a magnetic-field based energy-free, non-contact and non-invasive way.[16-18] These features consequently permit the fabrication of giant magneto-optic Cotton-Mouton (CM) devices capable of light manipulation[4, 18, 19] and the bionic/robotic material with anisotropic mechanic/electric properties.[20, 21] However, the collective behaviors of 2D colloidal particles have not been introduced and its fundamental role remains unclarified. Principally, a 2D colloidal system is a charge-stabilized dispersion, in which colloidal particle is surface-charged due to dissociation of the surface groups in water, forming electric double layers (EDLs) around the colloids. The ionic composition in the EDLs determines the Debye length of the system,[22] namely, the longest distance that the electrostatic force of the colloidal particle can exert. Therefore, modulation of ionic strength in the dispersion may provide an effective way to control the collective behavior and the consequent performance of 2D colloid based magneto-optic device.

Here we demonstrate an extremely large magneto-optic CM coefficient by introducing the collective behavior into 2D material dispersions. By adding ions into dispersion, the Debye length of the 2D cobalt-doped titanium oxide (CTO) colloid decreases, which strengthens inter-particle van der Waals attraction and promotes the formation of microdomains as evidenced by the small-angle X-ray scattering (SAXS) measurements. Each domain includes cofacially aligned CTO flakes and serves as a whole in response to the external magnetic stimulus. The CM coefficient consequently doubles and increases up to 2700 $T^{-2}m^{-1}$, which is the highest value obtained so far, permitting the observation of transmissive magneto-coloration in a low magnetic field about 0.1 T. Moreover, the peak wavelength of transmittance spectra of 2D CTO LC redshifts linearly with the increase of concentration and the decline of hydration radius of ions, allowing the finely engineerable magneto-coloration and the use of 2D CTO based inorganic LC devices as visual ion detectors.

**Results and discussion**

2D CTO with a lepidocrocite-type structure (Figure 1a) was prepared by high temperature solid state chemical reaction followed by exfoliation and was dispersed in water (for details see section 1 of the Supporting Information). The average thickness and lateral size of the CTO are 1.9 nm and 680 nm (Figure S1), and a typical CTO flake is shown in Figure 1b. All dispersions without a special note in this work have the same CTO concentration of 0.02 vol%. Moreover, the dispersions have an isotropic phase in the static state (left panel in Figure 1c) and a nematic phase during shaking (right panel in Figure 1c) as evidenced by the black and rainbow interference color in the presence of crossed polarizers, respectively. To investigate the impact of ions on the magneto-optic response of 2D CTO LC, two types of CTO LCs were prepared and labelled as "I-CTO" and "C-CTO" LCs. Here I-CTO LC denotes the individual-behavior dominated LC, which is the as-prepared CTO after centrifugation, while C-CTO stands for the collective-behavior dominated LC after intentional addition of sodium ions in the as-prepared CTO dispersion (for details see section 1 of the Supporting Information).

The optical setup for the magneto-coloration measurement is shown in Figure 1d, where two quartz cuvettes loaded with I-CTO and C-CTO LCs are placed between crossed polarizers with an angle of 45° between their transmission axis and magnetic flux. In the presence of an external magnetic field, the orientation preference of 2D CTO is determined by the magnetic torque acting on it. The negative magnetic susceptibility anisotropy of CTO forces these flakes to align parallel with the magnetic flux, which gives the final magneto-birefringence $\Delta n(H)$.[18, 23] For a medium with a thickness $L$, if the phase retardation of $\Delta n(H)L$ increases above 250 nm, the first-order yellow would emerge and undergo the color evolution from the first-order to the higher-order ones, according to the Michel-Levy chart.[24] Figure 1e shows that the addition of ions induces a distinct color shift from turquoise (greenish-blue) for I-CTO LC to cream (greenish-yellow) for C-CTO LC at a fixed magnetic strength of 0.45 T. Such ion-color dependence can be further identified by the redshift of peak wavelength in the transmittance spectra with the increase of ionic strength (Figure 1f).

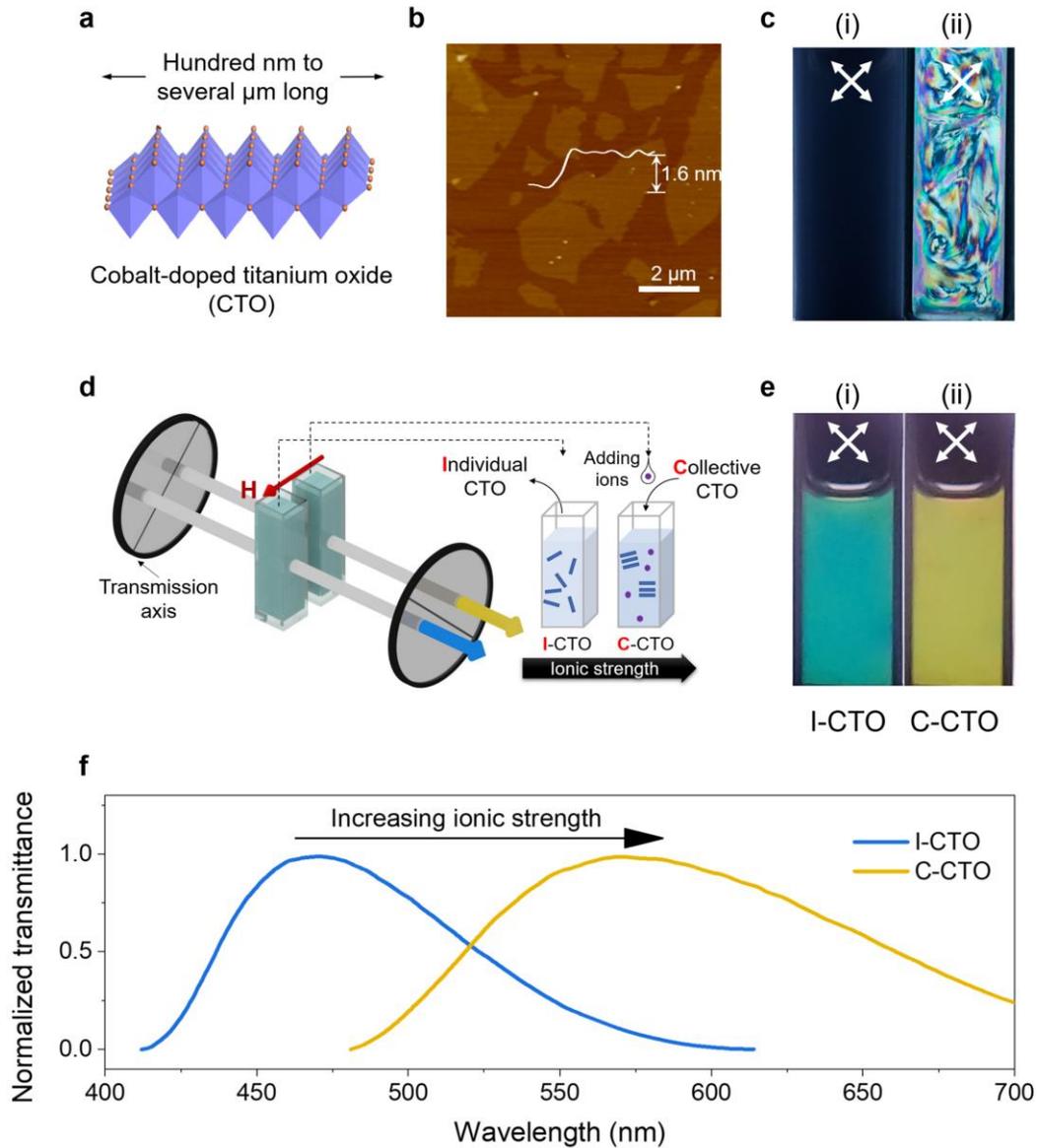

**Figure 1.** Magneto-optic effect and magneto-coloration of 2D CTO inorganic LCs with different ionic strengths. (a) Atomic structures of 2D CTO. (b) Atomic force microscopy (AFM) image of 2D CTO flakes deposited on a silicon wafer. (c) Polarized optical images of 2D CTO LC without (i) and with (ii) shaking. (d) Schematic setup of our magneto-optical experiments. The inset shows the illustration of the magnetic field induced alignment of 2D flakes in individual-behavior dominated CTO (I-CTO) and collective-behavior dominated CTO (C-CTO) LCs, corresponding to LCs without and with ion addition, respectively. (e) Polarized optical images of tunable interference colors of I-CTO (i) and C-CTO (ii) LCs under 0.45 T. (f) Normalized transmission spectra of I-CTO and C-CTO LCs at 0.45 T. The redshift of the peak wavelength results
5

from the ion addition.

To quantify the influence of ion addition on the magneto-coloration of 2D CTO, the field-color correspondence is recorded (Figure 2a). Taking I-CTO LC as an example, it appears as the first-order orange color in the magnetic field of 0.4 T. Whereafter, the color enters the second-order one and ends at fluro-pink at 0.8 T. Such field-color correspondence becomes more sensitive for C-CTO LC as evidenced by the reduced field strength to observe the same color. Transmittance mappings of I-CTO and C-CTO LCs, i.e., transmittance as a function of both wavelength $\lambda$ and magnetic field $H$, are shown in Figure 2b and 2c. Upon increasing the magnetic field from 0 to 0.8 T, the transmittance exhibits alternating oscillations of maxima (red stripes) and minima (purple stripes). The behavior can be well described by phase-retardation based Malus' law.[25] When the phase retardation of $\delta$ satisfies $\delta = \frac{2\pi \Delta n(H) L}{\lambda} = (2N-1)\pi$ or $(2N-2)\pi$, constructive/destructive interference dominates the maximum/minimum transmittances, where $N$ is the $N^{th}$ order maxima or minima. It is worth noting that without ions, the number $N$ of maximum stripes is 2 (in I-CTO), whereas, the addition of salt doubles $N$ to 4 (in C-CTO). In the condition of identical thickness $L$ for both LCs, the difference in $N$ indicates the increase in magneto-birefringence $\Delta n(H)$ by ion addition. The chromaticity diagrams (Figure 2d and 2e) further support the relationship. The color evolution of C-CTO LC covers the widest range and forms a closed heart-shaped loop in a small range of 0.125 to 0.45 T. The result is consistent with the observation in Figure 2a, where C-CTO LC requires a lower magnetic field than I-CTO does to display the same color. These results together confirm that the sensitivity of magneto-optic response is increased by ion addition and indicate tunability of magneto-coloration by controlling the ionic strength in the dispersion.

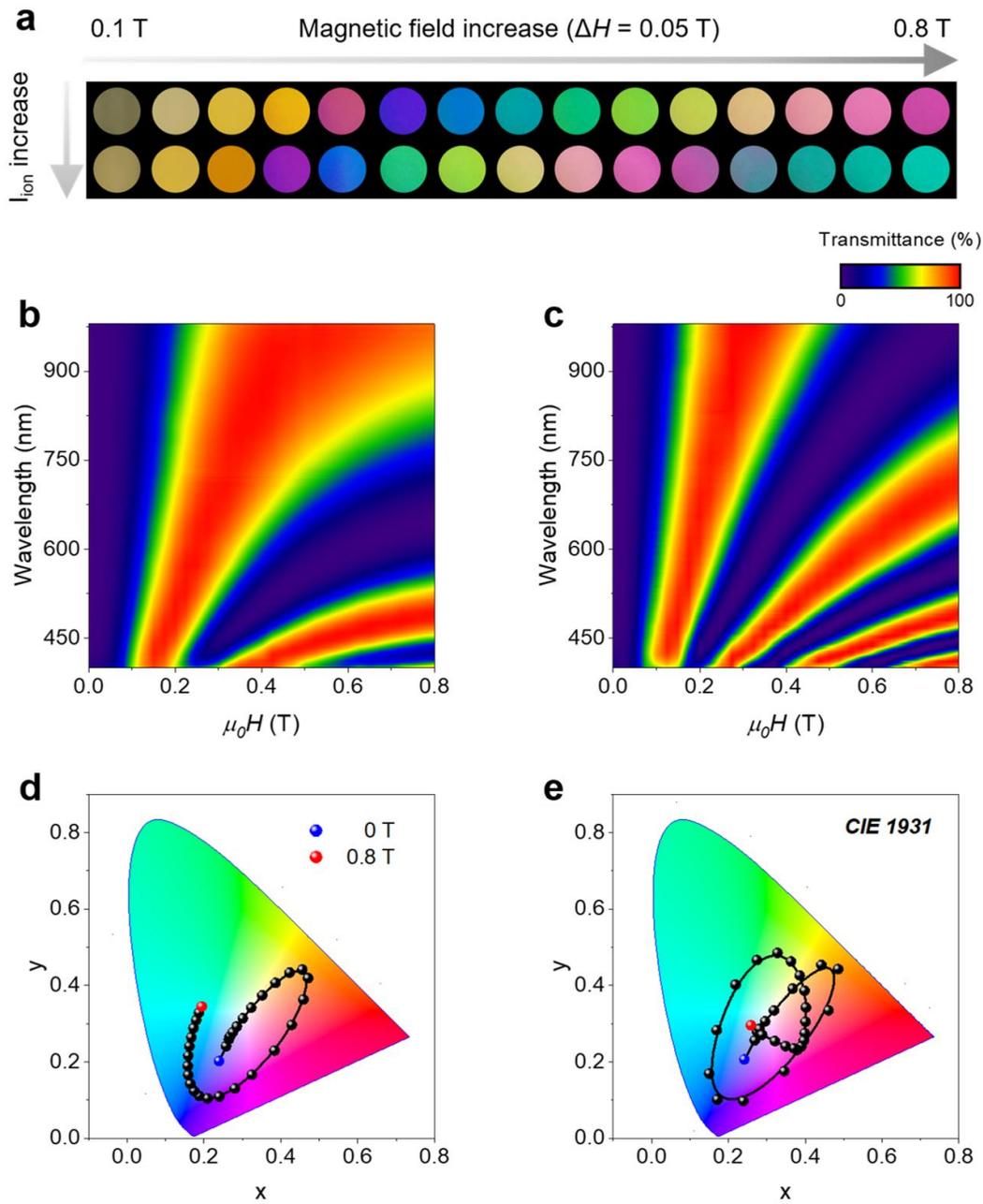

**Figure 2.** Optical characterization of 2D CTO inorganic LCs with different ionic strengths. (a) Snapshots of I-CTO and C-CTO LCs in the magnetic field range of 0.1 to 0.8 T with an interval of 0.05 T. (b,c) Transmittance mappings of I-CTO LC (b) and C-CTO LC (c) in the magnetic field range of 0 to 0.8 T. (d,e) Color evolutions of I-CTO LC (d) and C-CTO LC (e) presented in the CIE 1931 color space chromaticity coordinates.

To reveal the mechanism that dominates the ion-color correspondence in the given magnetic field, we investigated the influence of ion addition on the inter-particle interaction between 2D CTO materials. Figure 3a illustrates the formation of microdomains by ion addition. Principally, a 2D CTO flake is a charged colloidal particle and surrounded by an EDL. The addition of ions can induce the thinning of the EDL, attributed to the adsorption of added cations on the surface of CTO, which consequently decreases the Debye length $\kappa^{-1}$ according to the equation $\kappa^{-1} = \left(\frac{\varepsilon_0 \varepsilon_r k_B T}{2 N_A I e^2}\right)^{1/2}$, where $I$ the ion concentration, $\varepsilon_0$ the permittivity of vacuum, $\varepsilon_r$ the relative permittivity of water, $k_B$ the Boltzmann constant, $T$ the temperature, $N_A$ the Avogadro constant and $e$ the charge of an electron.[2] The Debye length determines the longest distance that the electrostatic force of the colloidal particle can exert. Therefore, upon decreasing $\kappa^{-1}$ by adding ions, the balance between long-range van der Waals attraction and electrostatic repulsion among CTO flakes will occur at a shorter inter-particle distance, which would possibly promote the formation of microdomains as schemed in Figure 3a. To test this, 1D small-angle X-ray scattering (SAXS) profiles (Figure 3b) of I-CTO and C-CTO LCs were measured by using synchrotron X-rays. A SAXS peak is observable at q = 0.031 Å$^{-1}$ for C-CTO LC, corresponding to an average inter-particle spacing of 20 nm, while no peak was detected in the whole range of 0.01 to 0.05 Å$^{-1}$ for I-CTO. We also calculated the secondary minimum of CTO surface potential by using Derjaguin–Landau–Verwey–Overbeek (DLVO) theory (for details see section 4 of the Supporting Information),[2] giving the inter-particle distances of 22 nm for C-CTO (Figure 3c), which is consistent with the experimental observation. We note that the I-CTO or C-CTO LC sample for SAXS measurements has a concentration of 0.12 vol%. The spacing between mutually independent CTO flakes for I-CTO can be calculated to be about 166 nm (Figure 3c) and is about eight-time larger than the value for C-CTO. In this regard, such small spacing for C-CTO indicates the formation of microdomains, each of which contains pieces of orderly aligned CTO flakes.

Moreover, the formation of domains can either decrease the aspect ratio of the functional unit (lateral size/thickness) or orientational entropy in the nematic phase, which consequently gives rise to the increase in phase-transition concentration of aqueous colloidal dispersions according to the Onsager theory[26], such as isotropic-to-biphasic one $\Phi_{I-B}$ and biphasic-to-nematic one $\Phi_{B-N}$ (for discussion in detail see section 5 of the Supporting Information). Figure 3d shows the images of two CTO dispersions in capillaries taken by a polarized optical microscope, through which a three-phase sequence can be identified. By measuring the volume fraction of the isotropic phase relative to the nematic phase, the phase diagram of the colloidal dispersion was determined (Figure 3e), giving $\Phi_{I-B}$ of 0.024 and 0.06 vol% for I-CTO and C-CTO, respectively. Both the maximum $\Phi_{I-B}$ and $\Phi_{B-N}$ are observed for the C-CTO LC, agreeing with the inference about increased phase-transition concentration by the ion-induced domain.

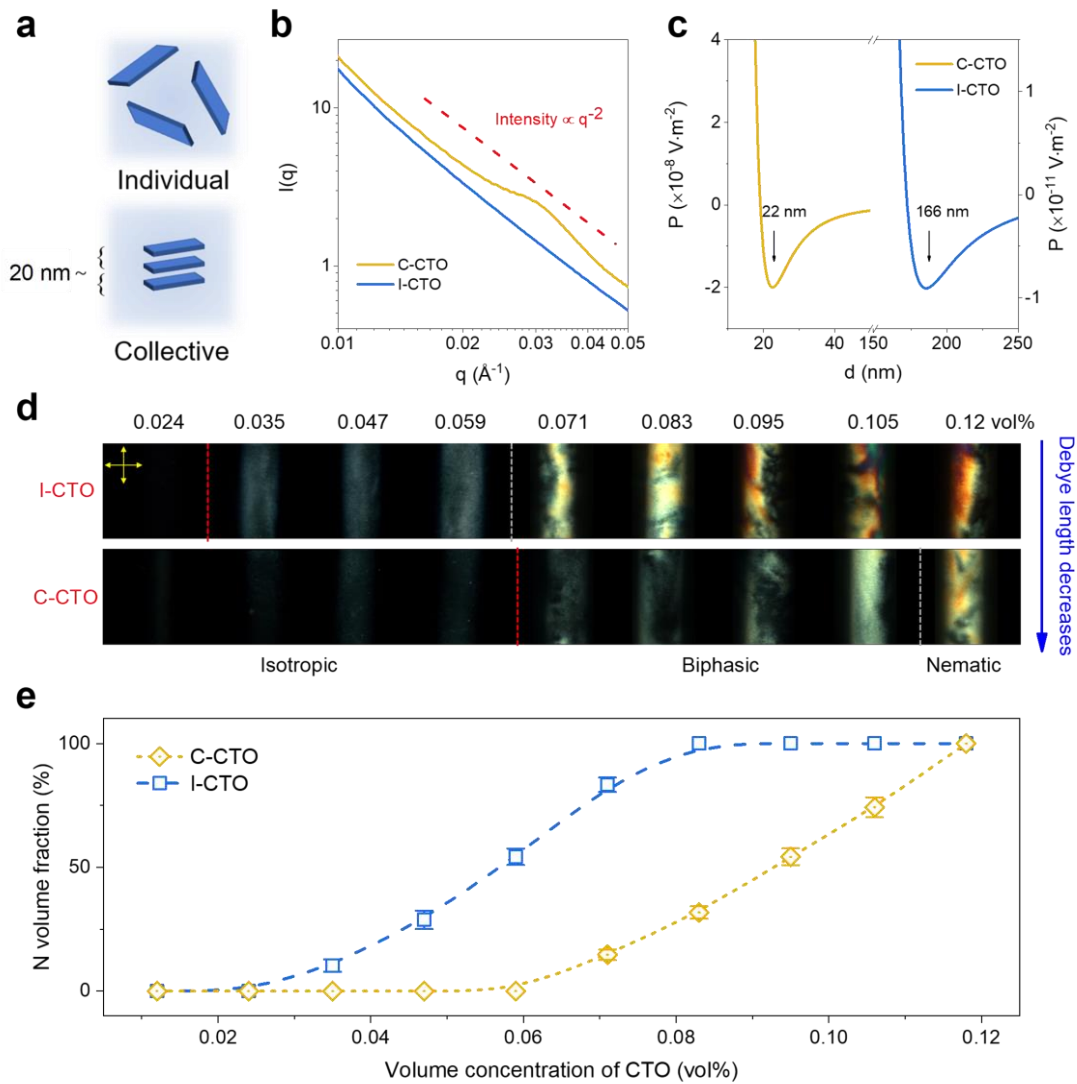

**Figure 3.** Collective behavior induced domain formation in 2D CTO. (a) Schematic of microdomain formation. The increase of ionic strength reduces the electrostatic repulsion between CTO flakes. The re-balance between weakened inter-particle electrostatic repulsion and strengthened van der Waals attraction results in the domain formation with an inter-particle distance about 20 nm. (b) 1D SAXS profiles of CTO aqueous dispersions with and without ions. (c) Theoretical interaction potential between a pair of I-CTO (right) and C-CTO (left) flakes calculated by the DLVO theory. (d) Polarized optical microscope images of capillaries containing CTO LCs with different ionic strengths. Isotropic-biphasic and biphasic-nematic transitions are marked with red and grey dashed lines, respectively. (e) Phase transition diagram of CTO LCs with different ionic strengths. The volume fraction is calculated by dividing the volume of the low-concentrated isotropic or biphasic phase with the high-concentration nematic phase.

For CTO flakes within a domain, they align cofacially and behave collectively, thereby being treated as a whole functional unit in response to the external magnetic stimulus. In the presence of a magnetic field, when the magnetic anisotropic energy of a functional unit is high enough, the influence of thermal disturbance is suppressed, resulting in the alignment of the unit along the external field. Such magnetic anisotropic energy is dominated by the number of cobalt atoms in the functional unit.[27] For I-CTO LC, the functional unit is made of individual CTO flake. In the given strength of the magnetic field, only large CTO flakes containing enough cobalt atoms can be aligned as schemed in Figure 4a. While for C-CTO LC, the functional unit of the microdomain has more cobalt atoms in comparison with that in I-CTO. Therefore, as illustrated in Figure 4b, more CTO flakes are aligned orderly. Finally, alignment and optical anisotropy of functional units collectively produce the magneto-birefringence $\Delta n(H)$ which can be expressed as $\Delta n(H) = \Delta n_P \phi S(H)$ where $\Delta n_P$ the specific birefringence, $\phi$ the volume fraction and $S(H)$ magnetic-field induced alignment order.[28, 29]

The increased magnetic sensitivity by the ion-induced domains is embodied in the larger magneto-optic CM coefficient, namely higher $S(H)$ and the corresponded $\Delta n$ at the same $H$ (Figure S4). As shown in Figure 4c, in the field range of 0 to 0.25 T, $\Delta n$ exhibits a linear increase with $H^2$, indicating a typical Cotton-Mouton effect as described by $\Delta n = C_{CM}\lambda H^2$, where $C_{CM}$ is the CM coefficient.[30] By calculating the slopes at $\lambda$ of 450 nm, $C_{CM}$ yields 1450 T$^{-2}$m$^{-1}$ for I-CTO LCs, which is consistent with the previously reported highest value of 1400 T$^{-2}$m$^{-1}$ for the similar system.[18] Upon raising ionic strength for C-CTO, the formed microdomains increase the coefficient up to 2700 T$^{-2}$m$^{-1}$. which is about twice the highest reported value and three to ten orders of magnitude larger than those of other known transparent media (Figure 4d). For example, these media include different types like 1) gas of helium, neon, argon, xenon; 2) pure solvent of acetone, benzene, acetonitrile, etc. 3) organic liquid crystal of 5CB, 5OCB, MBBA, etc. and 4) lyotropic liquid crystals of CTO. The obtained extremely large CM coefficient of C-CTO allows the observation of magneto-coloration in a low magnetic field of 0.125 T (Figure S5), which can be supplied by the energy-free permanent magnet instead of expensive superconducting- or electro-magnet.

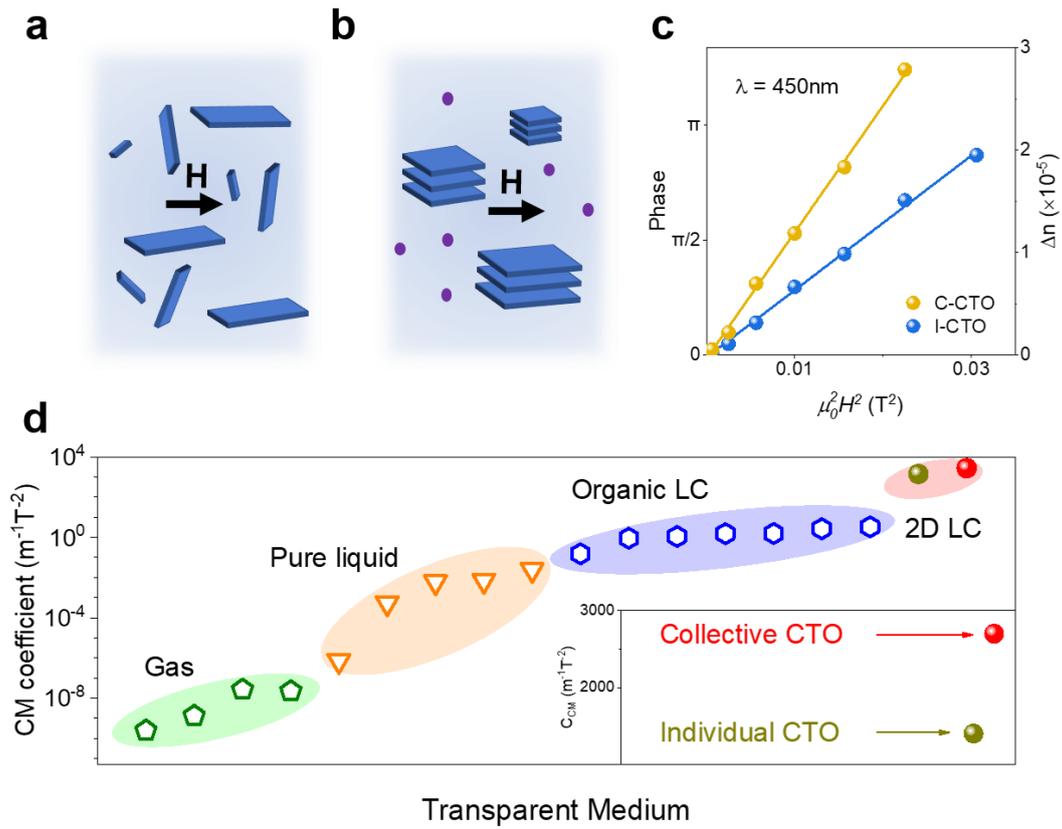

**Figure 4.** Enhanced Cotton-Mouton (CM) effect of 2D CTO LC. (a,b) Schematics of the alignment of I-CTO LCs (a) and C-CTO LCs with microdomain structure (b). (c) The phase retardation and birefringence presented as a function of the square of the magnetic strength in low field. The slope represents the CM coefficient. (d) A summary of the CM coefficient obtained so far among all transparent media.

Due to the ion-dependent domain formation, the magneto-coloration can be engineered by ion control. To examine the effect of ion concentration on the magneto-coloration, we obtained C-CTO LCs with ion concentrations of 0.2, 0.4, 0.6, 0.8 and 1 mM. Magneto-color evolves from green to orange (Figure S6) at 0.5 T, which can be quantitatively identified by the ion-dependent transmittance spectra (Figure 5a). The peak wavelength in the spectrum shows a linear response to the ionic strength in the dispersion, the slope of which gives the response sensitivity of 113 nm·mmol$^{-1}$ for sodium ions (Figure 5b). In the meantime, we also found that the response sensitivity is selectively dependent on the ion type. For example, under the same magnetic field of 0.5 T, the response sensitivity for the ions in the group IA are 65, 110, 148 and 162 nm·mM$^{-1}$ for Li$^+$, Na$^+$, K$^+$ and Cs$^+$, respectively (Figure 5c). By calculating their hydrated ion radii, we found a negative linear correlation between the response sensitivity with their hydrated ion radii,[31] while its slope gives the value of -1810 mM$^{-1}$ (Figure 5d). Based on the extremely large CM coefficient and the selectivity of magneto-coloration on both ion concentration and ion type, the CTO LC device can serve as the ion sensor with the advantages of portability, visualization and energy-free. As a concept demonstration, Figure 5e shows a visualized ion sensor, comprising a cuvette with CTO LC placed between two crossed polarizers in the light direction and two thumb-size permanent magnets in the left-and-right direction. By adding salt of CsCl, the uniform blue-purple color was disturbed and became random rainbow stripes. After one-minute stabilization, the color appeared as the uniform turquoise (blue-green) one (see supplementary video 1).

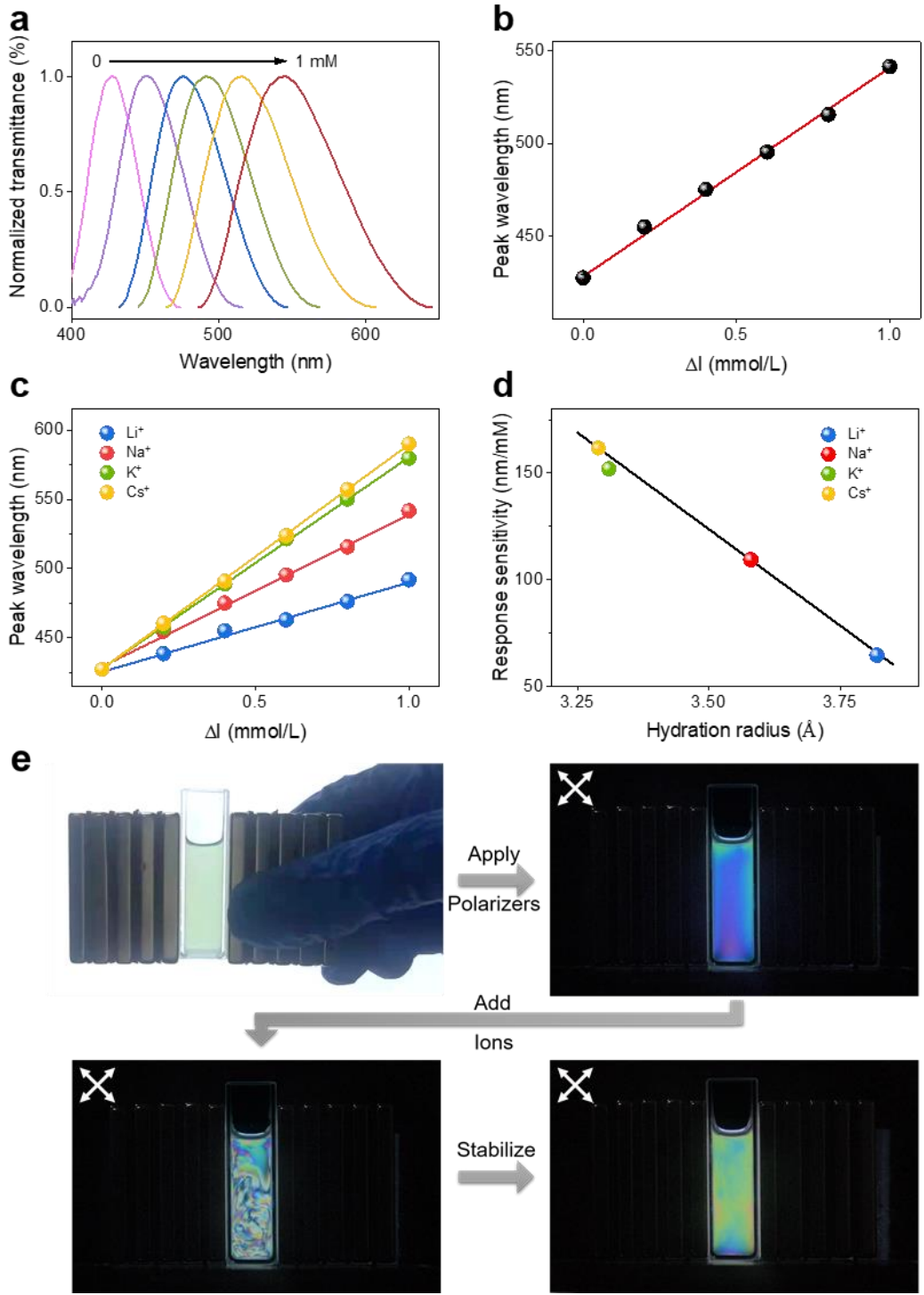

**Figure 5.** Dependence of magneto-coloration of 2D CTO on the concentrations and types of ions. (a,b) Normalized transmission spectra (a) and changes in peak wavelength (b) of CTO LC with different ion concentrations. (c) Peak wavelength changes in response to different ion species. The slope represents the response sensitivity of ions. (d) Response sensitivity verse hydration radius of ions. (e) Proof-of-concept portable, power-free and visual ion detector based on giant Cotton-Mouton effect of 2D CTO LC.

**Conclusion**

We have demonstrated the engineerable magneto-coloration of 2D CTO inorganic LC by controlling the formation of microdomains in the colloidal dispersion. The addition of ions reduces the Debye length of CTO flake and consequently induces the formation of microdomain with an inter-particle distance of 20 nm. The collective behavior of CTO flakes inside a domain makes them serve as a whole functional unit and shows higher sensitivity in response to the external magnetic stimulus than individual flakes. Such collective behavior leads to an extremely large magneto-optic Cotton-Mouton coefficient of 2700 $T^{-2}m^{-1}$ for CTO LC. The magneto-coloration of CTO LC shows a linear response to both the concentration and hydration radius of ions. Our work kindles a spark in revealing the fundamental role of collective behavior of colloidal particles in the stimuli-induced effects of colloidal system or lyotropic LC.

**Acknowledgement**

We acknowledge support by the National Natural Science Foundation of China (No. 51920105002), the Guangdong Innovative and Entrepreneurial Research Team Program (No. 2017ZT07C341), the Shenzhen Basic Research Project (Nos. JCYJ20190809180605522, JCYJ20200109144620815 and JCYJ20200109144616617), the National Key R&D Program (2018YFA0307200), and the Bureau of Industry and Information Technology of Shenzhen for the "2017 Graphene Manufacturing Innovation Center Project" (No. 201901171523). We also acknowledge BL16B1 beamline of Shanghai Synchrotron Radiation Facility for the trail test.

**Reference**

(1)  Woltman, S. J.; Jay, G. D.; Crawford, G. P. Liquid-Crystal Materials Find a New Order in Biomedical Applications. *Nat. Mater.* **2007**, *6*, 929-38.

(2) Sano, K.; Kim, Y. S.; Ishida, Y.; Ebina, Y.; Sasaki, T.; Hikima, T.; Aida, T. Photonic Water Dynamically Responsive to External Stimuli. *Nat. Commun.* **2016**, *7*, 12559.

(3) Shen, T. Z.; Hong, S. H.; Song, J. K. Electro-Optical Switching of Graphene Oxide Liquid Crystals with an Extremely Large Kerr Coefficient. *Nat. Mater.* **2014**, *13*, 394-9.

(4) Lin, F.; Zhu, Z.; Zhou, X.; Qiu, W.; Niu, C.; Hu, J.; Dahal, K.; Wang, Y.; Zhao, Z.; Ren, Z.; Litvinov, D.; Liu, Z.; Wang, Z. M.; Bao, J. Orientation Control of Graphene Flakes by Magnetic Field: Broad Device Applications of Macroscopically Aligned Graphene. *Adv. Mater.* **2017**, *29*, 1604453.

(5) Nakayama, M.; Kajiyama, S.; Kumamoto, A.; Nishimura, T.; Ikuhara, Y.; Yamato, M.; Kato, T. Stimuli-Responsive Hydroxyapatite Liquid Crystal with Macroscopically Controllable Ordering and Magneto-Optical Functions. *Nat. Commun.* **2018**, *9*, 568.

(6) Urbas, A.; Tondiglia, V.; Natarajan, L.; Sutherland, R.; Yu, H.; Li, J. H.; Bunning, T. Optically Switchable Liquid Crystal Photonic Structures. *J. Am. Chem. Soc.* **2004**, *126*, 13580-1.

(7) Luk, Y. Y.; Abbott, N. L. Surface-Driven Switching of Liquid Crystals Using Redox-Active Groups on Electrodes. *Science* **2003**, *301*, 623-6.

(8) White, T. J.; Broer, D. J. Programmable and Adaptive Mechanics with Liquid Crystal Polymer Networks and Elastomers. *Nat. Mater.* **2015**, *14*, 1087-98.

(9) Sagara, Y.; Kato, T. Mechanically Induced Luminescence Changes in Molecular Assemblies. *Nat. Chem.* **2009**, *1*, 605-10.

(10) Kato, T. Self-Assembly of Phase-Segregated Liquid Crystal Structures. *Science* **2002**, *295*, 2414-8.

(11) Palmer, L. C.; Leung, C. Y.; Kewalramani, S.; Kumthekar, R.; Newcomb, C. J.; Olvera de la Cruz, M.; Bedzyk, M. J.; Stupp, S. I. Long-Range Ordering of Highly Charged Self-Assembled Nanofilaments. *J. Am. Chem. Soc.* **2014**, *136*, 14377-80.

(12) Krause, J.; Ruxton, G. D.; Krause, S. Swarm Intelligence in Animals and Humans. *Trends Ecol. Evol.* **2010**, *25*, 28-34.

(13) Sasaki, T.; Biro, D. Cumulative Culture Can Emerge from Collective Intelligence in Animal Groups. *Nat. Commun.* **2017**, *8*, 15049.

(14) Krausz, R. R. Living in Groups. *Transactional Analysis Journal* **2017**, *43*, 366-374.

(15) Maciołek, A.; Dietrich, S. Collective Behavior of Colloids Due to Critical Casimir Interactions. *Rev. Mod. Phys.* **2018**, *90*, 045001.

(16) Faure, B.; Salazar-Alvarez, G.; Bergstrom, L. Hamaker Constants of Iron Oxide Nanoparticles. *Langmuir* **2011**, *27*, 8659-64.

(17) Anandarajah, A.; Lu, N. Numerical Study of the Electrical Double-Layer Repulsion between Non-Parallel Clay Particles of Finite Length. *Int. J. Numer. Anal. Methods Geomech.* **1991**, *15*, 683-703.

(18) Ding, B.; Kuang, W.; Pan, Y.; Grigorieva, I. V.; Geim, A. K.; Liu, B.; Cheng, H. M. Giant Magneto-Birefringence Effect and Tuneable Coloration of 2D Crystal Suspensions. *Nat. Commun.* **2020**, *11*, 3725.

(19) Kim, J. E.; Han, T. H.; Lee, S. H.; Kim, J. Y.; Ahn, C. W.; Yun, J. M.; Kim, S. O. Graphene Oxide Liquid Crystals. *Angew. Chem., Int. Ed. Engl.* **2011**, *50*, 3043-7.

(20) Liu, M.; Ishida, Y.; Ebina, Y.; Sasaki, T.; Hikima, T.; Takata, M.; Aida, T. An Anisotropic Hydrogel with Electrostatic Repulsion between Cofacially Aligned Nanosheets. *Nature* **2015**, *517*, 68-72.

(21) Kim, Y. S.; Liu, M.; Ishida, Y.; Ebina, Y.; Osada, M.; Sasaki, T.; Hikima, T.; Takata, M.; Aida, T. Thermoresponsive Actuation Enabled by Permittivity Switching in an Electrostatically Anisotropic Hydrogel. *Nat. Mater.* **2015**, *14*, 1002-7.

(22) Bonn, D.; Otwinowski, J.; Sacanna, S.; Guo, H.; Wegdam, G.; Schall, P. Direct Observation of Colloidal Aggregation by Critical Casimir Forces. *Phys. Rev. Lett.* **2009**, *103*, 156101.

(23) Verhoeff, A. A.; Brand, R. P.; Lekkerkerker, H. N. W. Tuning the Birefringence of the Nematic Phase in Suspensions of Colloidal Gibbsite Platelets. *Mol. Phys.* **2011**, *109*, 1363-1371.

(24) Sørensen, B. E. A Revised Michel-Lévy Interference Color Chart Based on First-Principles Calculations. *Eur. J. Mineral.* **2013**, *25*, 5-10.

(25) Wang, M.; He, L.; Zorba, S.; Yin, Y. Magnetically Actuated Liquid Crystals. *Nano Lett.* **2014**, *14*, 3966-71.

(26) Onsager, L. The Effects of Shape on the Interaction of Colloidal Particles. *Ann. N. Y. Acad. Sci.* **1949**, *51*, 627-659.

(27) Nair, R. R.; Sepioni, M.; Tsai, I. L.; Lehtinen, O.; Keinonen, J.; Krasheninnikov, A. V.; Thomson, T.; Geim, A. K.; Grigorieva, I. V. Spin-Half Paramagnetism in Graphene Induced by Point Defects. *Nat. Phys.* **2012**, *8*, 199-202.

(28) Lemaire, B. J.; Davidson, P.; Ferre, J.; Jamet, J. P.; Petermann, D.; Panine, P.; Dozov, I.; Jolivet, J. P. Physical Properties of Aqueous Suspensions of Goethite (Alpha-Feooh) Nanorods. Part I: In the Isotropic Phase. *Eur. Phys. J. E* **2004**, *13*, 291-308.

(29) Dozov, I.; Paineau, E.; Davidson, P.; Antonova, K.; Baravian, C.; Bihannic, I.; Michot, L. J. Electric-Field-Induced Perfect Anti-Nematic Order in Isotropic Aqueous Suspensions of a Natural Beidellite Clay. *J. Phys. Chem. B* **2011**, *115*, 7751-65.

(30) Adamczyk, A. Cotton-Mouton Effect in Monodisperse Suspensions of Liquid-Crystals. *Mol. Cryst. Liq. Cryst.* **1989**, *167*, 7-26.

(31) Nightingale, E. R. Phenomenological Theory of Ion Solvation - Effective Radii of Hydrated Ions. *J. Phys. Chem.* **1959**, *63*, 1381-1387.

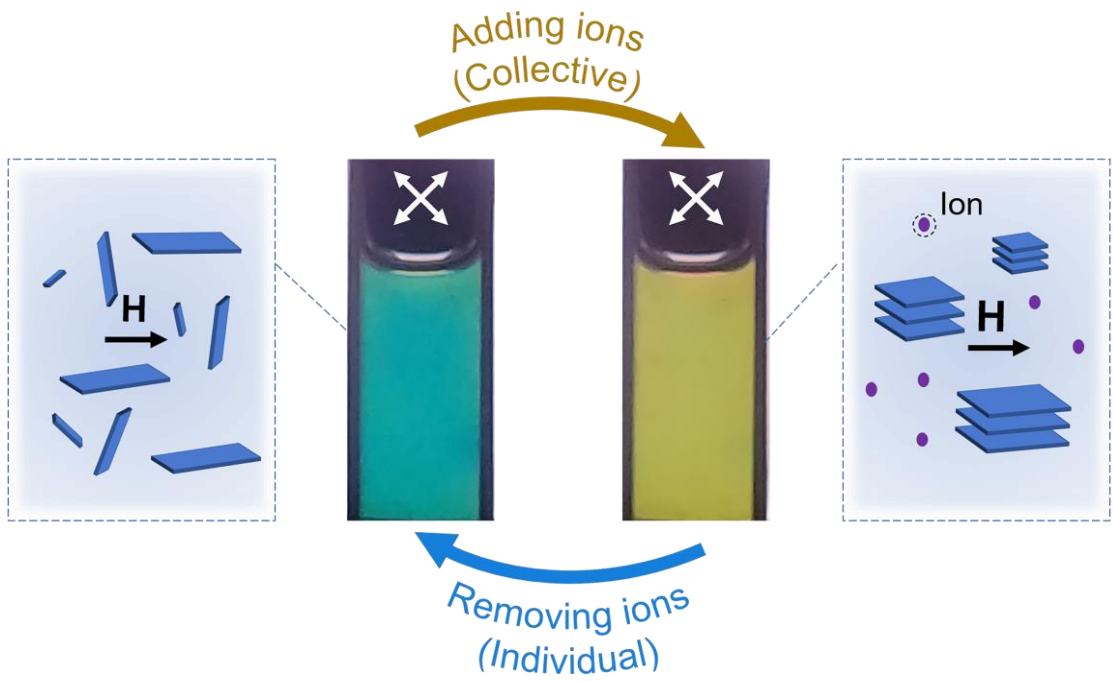

**For Table of Contents Only**